\documentclass[10pt,conference]{IEEEtran}
\usepackage{cite}
\usepackage{amsmath,amssymb,amsfonts}
\usepackage{mathtools, bigstrut}
\usepackage{multirow}
\usepackage{cuted}
\usepackage{booktabs}

\usepackage{lipsum}
\usepackage{algorithm}
\usepackage{algpseudocode}
\usepackage{physics}
\usepackage{caption}
\usepackage{subfig}
\usepackage{tikz}
\usepackage{algpseudocode}
\usepackage{graphicx}
\usepackage{textcomp}
\usepackage{xcolor}
\usepackage{hyperref}

\makeatletter 
\let\old@ps@headings\ps@headings 
\let\old@ps@IEEEtitlepagestyle\ps@IEEEtitlepagestyle 
\def\confheader#1{%
\def\ps@headings{%
\old@ps@headings%
\def\@oddhead{\strut\hfill#1\hfill\strut}%
\def\@evenhead{\strut\hfill#1\hfill\strut}%
}%
\def\ps@IEEEtitlepagestyle{%
\old@ps@IEEEtitlepagestyle%
\def\@oddhead{\strut\hfill#1\hfill\strut}%
\def\@evenhead{\strut\hfill#1\hfill\strut}%
}%
\ps@headings%
} 
\makeatother 

\confheader{}

\begin{document}

\title{Parametric entropy based Cluster Centriod Initialization  for $k$-means clustering of various Image datasets}

\author{\IEEEauthorblockN{Faheem Hussayn\IEEEauthorrefmark{1},~Shahid M Shah \IEEEauthorrefmark{2}}
\IEEEauthorblockA{Communication Control \& Learning Lab ($C^2L^2$)\\Department of Electronics \& Communication Engineering,\\National Institute of Technology Srinagar\IEEEauthorrefmark{2}\IEEEauthorrefmark{1}
\\
Email: faheemhussayn@gmail.com\IEEEauthorrefmark{1}, shahidshah@nitsri.ac.in\IEEEauthorrefmark{2}}

}

\IEEEoverridecommandlockouts 
\IEEEpubid{\makebox[\columnwidth]{} 
\hspace{\columnsep}\makebox[\columnwidth]{ }} 

\maketitle

\begin{abstract}
One of the most employed yet simple algorithm for cluster analysis is the $k$-means algorithm. $k$-means has successfully witnessed its use in artificial intelligence, market segmentation, fraud detection, data mining, psychology, etc., only to name a few. The $k$-means algorithm, however, does not always yield the best quality results. Its performance heavily depends upon the number of clusters supplied and the proper initialization of the cluster centroids or seeds. \\
In this paper, we conduct an analysis of the performance of $k$-means on image data by employing parametric entropies in an entropy based centroid initialization method and propose the best fitting entropy measures for general image datasets. We use several entropies like Taneja entropy, Kapur entropy, Aczel Daroczy entropy, Sharma Mittal entropy. We observe that for different datasets, different entropies provide better results than the conventional methods. We have applied our proposed algorithm on these datasets: Satellite, Toys, Fruits, Cars, Brain MRI, Covid X-Ray.

\end{abstract}


\section{Introduction}
A subset of artificial intelligence, machine learning, supplies machines with the ability to learn and make decisions without needing to be programmed explicitly. If learning is accomplished without supplying a machine with a labeled dataset, the machine needs to find the implicit data patterns without external support, this is known as unsupervised machine learning. Many real-life problems are actually modeled this way because, in real life, the pattern of data is difficult to know in advance \cite{ford2018architects}. \\
Recent times have witnessed an avalanche of data. The use of data mining techniques has thus seen a tremendous increase and clustering has been one of the most used unsupervised techniques. Clustering, sometimes referred to as cluster analysis, is an unsupervised machine learning technique wherein we tackle the problem of grouping or division of data points such that those data points that fall within the same group are more related to each other and less related to the data-points clustered into other groups \cite{jain1988algorithms}. It can also be simply defined as the collection or grouping of objects on the basis of similarity and dissimilarity between the objects.  This technique finds its applications in Wireless networks, System diagnostics, Search engines, Fraud detection, Market Segmentation, Satellite imagery, pattern recognition, big data analytics, and so on.
\\The simplest unsupervised learning algorithm that is usually employed in solving clustering problems is the $k$-means algorithm. $k$-means, being one of the most famous algorithms employed for clustering, has also witnessed its use as part of other algorithms \cite{saxena2017review}. The $k$-means algorithm is iterative in nature and aims to assign every data-point of a data-set to one of the $k$ clusters based on the features supplied. The $k$-means algorithm partitions '$n$' data points or observations into ‘$k$’ groups or clusters. $K$-means algorithm has applications in various areas, like energy analytics \cite{shah2021modelling}, attack detection \cite{ahmad2021mitigating}, \cite{ahmad2022supervised}, \cite{shahnawaz2022unsupervised}. Very recently $k$-means algorithm has also been used in activity detection in smart grid-based systems \cite{nida2022}. However, the quality of the solution and convergence speed of the $k$-means algorithm largely depends on the number of clusters supplied and the position of the initial seed points or cluster centroids. The traditional $k$-means algorithm initializes cluster centroids randomly but this has obvious drawbacks. Various methods have been devised for alternate centroid initialization, which will be discussed in the next section. In this study, we will focus on an centroid initialization method based on the maximization of entropy measure.\\


\section{Related Works}
For the problem of proper centroid initialization, two groups of studies exist. The first group of studies has focused on improving the existing random initialization. When incorporating entropy, Steinbach et al. found that the "bisecting $k$-means" method generally outperformed the classical $k$-means and when not measuring entropy, it almost performed similarly \cite{steinbach2000comparison}. However, they did not include any time-related metrics for comparison. $k$-means++ is another such modification of random centroid initialization. In this approach, Arthur and Vassilvitskii \cite{arthur2006k} initialized centroids from the data points at random while using the squared distance from the already initialized centroids to weigh potential centroids. The effect was that in contrast to the random initialization, this approach would ensure a maximal distance or "spread" of cluster centers or centroids.
\\The other group of studies has devised alternative methods to random centroid initialization for $k$-means and are consistent to the method we will discuss. A method for seed point selection that is recursive in nature was discussed by Duda and Hart \cite{duda1973pattern}. A MaxMin algorithm was devised by Higgs et al. \cite{higgs1997experimental} and Snarey et al. \cite{snarey1997comparison} which was based on selecting a subset from original database to be used for initial centroids in order to create initial clusters. The bilinear program was introduced by Bradley et al. \cite{bradley1998refining} which determined initial points in such a way that the sum of distances of each data point should be minimized to the nearest centroid. Su \& Dy \cite{su2004deterministic} came up with a deterministic method for centroid initialization. The method is hierarchically divisive in nature and is based on Principal Component Analysis (PCA). Cao et. al came up with an effective method to initialize clusters based on cohesion and coupling degree \cite{cao2009initialization}. Bai et. al proposed an initialization method based on distance and density metrics \cite{bai2012cluster}. Using the concept of Voronoi circles and their radii, Reddy et. al came up with an initialization technique \cite{reddy2012initialization}. Mahmud et al. came up with a method which is faster than traditional k-means. In this method, the selection of initial points is carried out using a weighted average score on the sorted data \cite{mahmud2012improvement}. A density based approach was developed by Gingles and Celebi \cite{gingles2014histogram} which was based on the hypothesis that centroids of clusters would naturally occur near the areas of high data-point density. Another density-based approach by Dalhatu et. al \cite{dalhatu2016density} has also been developed. One of the recent works for initialization based on entropy and with respect to image segmentation (clustering of pixels) was carried out by Chowdhury et al. \cite{chowdhury2018seed}. In their work, they employed the maximization of Shannon's Entropy to determine the optimal initial positions of the cluster centroids. This yielded lesser computation time and number of iterations with respect to image datasets. The given method, however, was only tested on select images using Shannon's entropy only. In spite of all these methods, presently, there is no universally accepted method for centroid initialization of $k$-means algorithm which is the prime reason for the pursuit of this study.

\section{Main Contribution}
The traditional $k$-means algorithm initializes centroids randomly and as already discussed, the quality of clustering depends upon the location of the initial centroids. We have employed the entropy maximization algorithm devised by Chowdhury et al. \cite{chowdhury2018seed} to initialize the centroids. However, in place of Shannon's Entropy, we test out different parametric entropy measures on contrasting image datasets with different parameters to yield the best fitting entropy for centroid initialization.
The entropy based initialization works on the entropy maximization principle. Exploiting the fact that for a multi spectral image, the intensity values for each color band of a particular pixel are mutually independent, we can easily calculate the probability of a pixel. Let N denote the total number of pixels in a given image and $a$, $b$, and $c$ be the intensity values of the Red, Green, and Blue color bands respectively. Also, let $n_a$, $n_b$ and $n_c$ be the number of intensity values for $a$, $b$ and $c$, respectively.
Then, using the concept of independent random variables we can arrive at the equation:
$$P(R=a,G=b,B=c)$$
$$P(R=a)*P(G=b)*P(B=c)$$
$$=\frac{n_a}{N}*\frac{n_b}{N}*\frac{n_c}{N}, \forall 0\leq a,b,c \leq 255$$
Using this probability, we can calculate the entropy measure for all the intensities in an image.

\section{Methodology and Datasets}
The algorithm for calculating the initial cluster centroids is given as Algorithm 1. For the entropy calculation step, we use the following entropy measures.\\
\textbf{Shannon Entropy}\\
Proposed by C.E Shannon \cite{shannon1948mathematical}, it is given as:\\
\begin{equation}
    H_S(P) = -\sum_{i=1}^np_i\log p_i
\end{equation}
\\
\textbf{Kapur}\\
Proposed by JN Kapur \cite{kapur1967generalized}, this entropy is given by:\\
\begin{equation}
   H_K(P) = \frac{1}{1-\alpha} \log\left(\frac{\sum_{i=1}^np_i^{\alpha+\beta-1}}{\sum_{i=1}^np_i^\beta}\right)
   \end{equation}
where $\alpha \neq 1, \alpha>0,\beta \geq 1$
\\
\\
\textbf{Aczél Daróczy}\\
Proposed by J Aczél, Z Daróczy \cite{aczel1963verallgemeinerte}, it can be calculated using:\\
\begin{equation}
H_{AD}(P) = \frac{1}{\beta} \arctan \left(\frac{\sum_{i=1}^np_i^\alpha \sin(\beta\log p_i)}{\sum_{i=1}^np_i^\alpha \cos(\beta\log p_i)}\right)
\end{equation}
where $\beta \neq 0, \alpha>0$
\\
\\
\textbf{Havrda and Charvát}\\
Proposed by J Havrda and F Charvát \cite{havrda1967quantification}, this entropy is given by:
\begin{equation}
   H_{HC}(P) = \frac{1}{(2^{1-\alpha}-1)} \left[\sum_{i=1}^np_i^\alpha-1\right]
   \end{equation}
where $\alpha \neq 1, \alpha>0$
\\
\\
\textbf{Taneja}\\
Proposed by I.J Taneja \cite{taneja1989generalized}, this entropy is calculated using the formula:\\
\begin{equation}
    H_{T}(P) = -\frac{2^{\alpha-1}}{\sin \beta}\sum_{i=1}^n p_i^\alpha\sin (\beta \log p_i)
\end{equation}
where $\alpha \neq k\pi,k=0,1,2,...,k>0$
\\
\\
\textbf{Sharma Mittal}\\
Proposed by Sharma and Mittal \cite{sharma1975new}, this entropy is given by:
\begin{equation}
    H_{SM}(P) =  \frac{1}{(2^{1-\alpha}-1)}\left[\left(\sum_{i=1}^np_i^\beta\right)^{\frac{\alpha-1}{\beta-1}}-1\right]
\end{equation}
where $\alpha \neq 1, \alpha>,\beta \neq 1, \beta>0$
\\
In this research, instead of using a few images to test the initialization, we have used different real life data-sets containing similar images and averaged out the results to evaluate time-related metrics for each entropy measure. The data-sets were obtained from publicly available sources of Kaggle and some of the images were manually curated from Google. A summary of the datasets used is given in Table 1. \\
Since in this study we have focused on the initialization method of the cluster centroids, we need a way to determine the optimal number of clusters as it is also a factor on which the quality of $k$-means clustering depends as already discussed. We used the classic `elbow method' to achieve this. The basic principle of the elbow method is that it plots the cost function (sum of square error values) for different values of $k$. Clearly, as the number of clusters increase, the SSE will reduce. A point will be reached where increasing the number of clusters will not have a drastic effect on the cost function. This we take as the optimal value of $k$.
We can have the sum of squared distances of all data-points to the cluster as the cost function, where we call it `inertia' or we can have the mean of squared distances of each data point to its nearest cluster, where we call it `dispersion'. \\
Since there are multiple images in a particular dataset, using the fact that the distribution of pixel intensities will be similar, we employed the elbow method on any one of the images in a particular image dataset to determine the optimal value for the number of clusters $k$ and conducted the analysis for all the images.  

\begin{table}[htbp]
  \centering
  \caption{Dataset Details}
    \begin{tabular}{lrrrl}
    \toprule
    Dataset & \multicolumn{1}{l}{Image Count} & \multicolumn{1}{l}{Attributes} & \multicolumn{1}{l}{Optimal k} & Source \\
    \midrule
    Satellite & 25    & 3     & 3     & Kaggle \\
    Toys  & 50    & 3     & 4     & Google \\
    Brain MRI & 30    & 2     & 3     & Kaggle \\
    X-Ray & 25    & 2     & 2     & Kaggle \\
    Fruits & 40    & 3     & 5     & Google \\
    Cars  & 50    & 3     & 3     & Kaggle \\
    \bottomrule
    \end{tabular}%
  \label{tab:addlabel}%
\end{table}%
An example of clustering using this approach is shown in Figure 1 and the corresponding comparison for the number of iterations utilized by $k$-means to converge is given in Figure 2. The $th$ value for this image is set to 220. 
\begin{figure}[H]
    \centering
    \includegraphics[scale=.5]{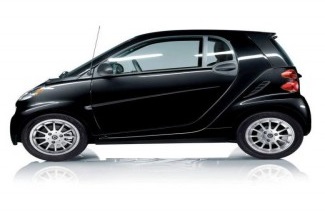}
    \label{fig:my_label}
    \caption{Original Image }
\end{figure}
\begin{figure}[H]
    \centering
    \includegraphics[scale=.9]{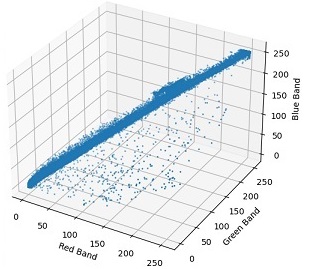}
    \label{fig:my_label}
    \caption{Scatter Plot}
\end{figure}
\begin{figure}[H]
    \centering
    \includegraphics[scale=.5]{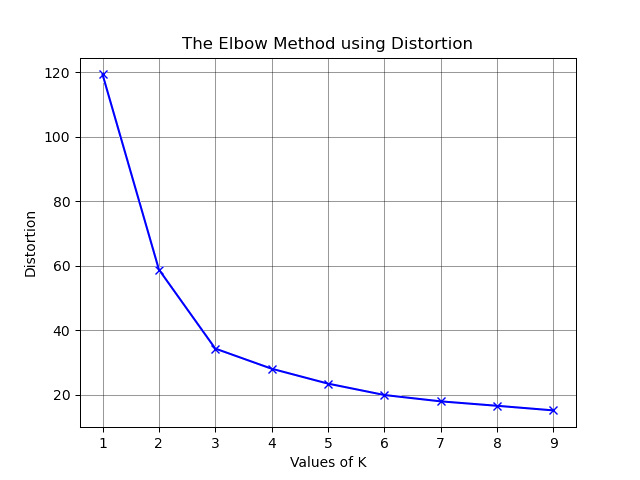}
    \label{fig:my_label}
    \caption{Distortion based Elbow Method}
\end{figure}
\begin{figure}[H]
    \centering
    \includegraphics[scale=.5]{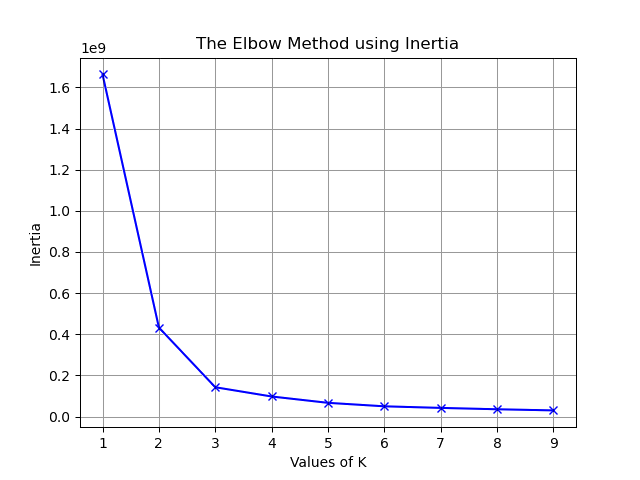}
    \label{fig:my_label}
    \caption{Inertia based Elbow Method}
\end{figure}
\begin{figure}[H]
    \centering
    \includegraphics[scale=.8]{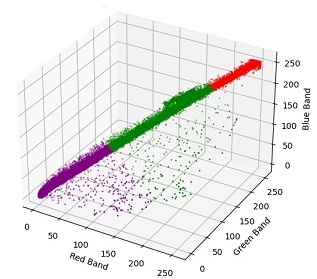}
    \label{fig:my_label}
    \caption{Optimal Clustering (K=3)}
\end{figure}
\begin{figure}[H]
    \centering
    \includegraphics[scale=.7]{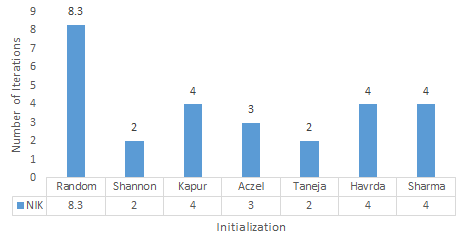}
    \caption{Iteration Comparison of Initializations for Car Image}
    \label{fig:my_label}
\end{figure}
It is clear from the above graph that the most appropriate initialization for clustering this image using $k$-means algorithm is the Shannon and Taneja Entropy. The random initialization denotes the default random initialization of the $k$-means algorithm. The image shown in Figure 1: (a) is derived from the car dataset. We cannot possibly show the clustering for each image used in the experiment as we have employed image datasets and not single images.

\algrenewcommand\algorithmicrequire{\textbf{Input:}}
\algrenewcommand\algorithmicensure{\textbf{Output:}}    
\begin{algorithm}
        \caption{Entropy Maximization Initialization}
        \label{alg:ALG1}
        \begin{algorithmic}
        \Require Image (dataset) and number of clusters ($K$)
        \Ensure Initial Centroid List
        \begin{enumerate}
    \item Input the number of cluster $K$ and $th$ (threshold for centroid spacing).
    \item Initialize the number of seeds $n_{cen}=1$
    \item Calculate needed entropy for each pixel in the image.
    \item Sort the pixels in descending order of entropy values.
    \item Take the first pixel from the sorted list and include in centroid list.
    \item Take the next pixel from the list and calculate its euclidean distance with all the pixels in SE.
    \item if $min_{ED}> TH$, include this pixel in centroid list and perform $n_{cen}=n_{cen}+1$. Otherwise goto step 6.
    \item if $n_{cen} = K$, stop. Otherwise goto step 6.
\end{enumerate}
        \end{algorithmic}
    \end{algorithm}
The $th$ or threshold value is a critical variable that essentially dictates the distance between the clusters. If it is not properly initialised, the distribution of centroids will not be appropriate. For example, if the value of $th$ is too large, the computation cost will rise and so will be the time taken for convergence. On the contrary, a value that is too small can cause the centroids to be very near causing the algorithm to converge prematurely. So, the choice of threshold value is important and and should be decided by considering the "spread" of the data.\\

\section{Results and Discussions}
In our experiment, we used several images from contrasting image data-sets to analyse the performance of multiple entropy measures on the cluster initialisation. The results were evaluated using the metrics: number of iterations of K-means (NIK), which essentially is the number centroid movements it takes to converge KMeans, Computation Time ($C_T$), which is the time to convergence, and Initialization Time ($I_T$). We club the Computation and Initialization time and call it Total Time. Since the $k$-means algorithm always converges and we are only modifying the initialization approach using different entropies, the sum of squared errors (SSE) metric may have similar results so our focus would be more on time-related metrics. The comparative description of the results is given in Table II. The initialization method using Shannon entropy is the original entropy maximization method devised by Chowdhury et al. \cite{chowdhury2018seed}. \cite{chowdhury2018seed}

\begin{table}[htbp]
  \centering
  \caption{Comparison of Initialization for Image Datasets}
    \begin{tabular}{rlrrr}
    \toprule
    \multicolumn{1}{l}{Dataset} & Initialization & \multicolumn{1}{l}{Avg. NIK} & \multicolumn{1}{l}{Total Time} & \multicolumn{1}{l}{SSE} \\
    \midrule
    \multicolumn{1}{l}{Satellite} & Random & 4.76  & 2.148 & 3751.72 \\
          & Shannon & \textbf{4.04} & \textbf{1.84} & \textbf{3751.5} \\
    \midrule
    \multicolumn{1}{l}{Toys} & Random & 4.07  & 1.4335 & 1493.13 \\
          & Shannon & 4.21  & 1.819 & \textbf{1492.3} \\
          & Taneja & \textbf{3.11} & \textbf{1.079} & \textbf{1492.3} \\
    \midrule
    \multicolumn{1}{l}{Fruits} & Random & 5.2   & 1.822 & 1564.1 \\
          & Shannon & 6.3   & 1.938 & 1564.6 \\
          & Taneja & \textbf{3.1} & \textbf{0.469} & \textbf{1563.6} \\
    \midrule
    \multicolumn{1}{l}{Cars} & Random & 3.9   & 0.9242 & 1331.91 \\
          & Shannon & 4.39  & 1.469 & 1332.02 \\
          & Taneja & \textbf{2.01} & \textbf{0.401} & \textbf{1330.32} \\
    \midrule
    \multicolumn{1}{l}{Brain MRI} & Random & 4.91  & \textbf{0.0203} & 1364.66 \\
          & Shannon & 6.285 & 0.065 & 1364.5 \\
          & Kapur & \textbf{4.285} & 0.036 & \textbf{1363.42} \\
    \midrule
    \multicolumn{1}{l}{Covid X-Ray} & Random & 4.89  & 0.05  & 1379 \\
          & Shannon & 3.69  & 0.079 & 1378.34 \\
          & Kapur & \textbf{2.24} & \textbf{0.038} & \textbf{1377.43} \\
    \end{tabular}%
  \label{tab:addlabel}%
\end{table}%
From our study we concluded that there was no single entropy that was appropriate for the cluster initialization of every kind of image dataset. We get the insight that for certain datasets, certain entropy measures worked better. 
We summarize our results with the following insights:
\begin{itemize}
    \item For the datasets with natural intensity levels and a higher dynamic range such as the images of cars, robots, toys, vegetables, fruits, etc. Taneja Entropy was the most appropriate.
    \item For the datasets with wide range of details like the satellite imagery, Shannon Entropy was the most appropriate for cluster initialization.
    \item For the datasets with similar saturation and less dynamic ranges, like the medical datasets of X-Ray and MRI Images, Kapur's Entropy was the most appropriate.
\end{itemize}

\section{Conclusion and Future Scope}
In our study, we have extended the entropy based initialization method for image clustering using $k$-means by employing parametric entropy measures and demonstrated its effectiveness on contrasting image datasets. To generalize the results found in this study and establish facts based upon them would be irresponsible at this point due to the small size and lesser number of the datasets used. However, our findings do point out that further research should be pursued, and that the further exploration of the parametric entropies discussed in the previous sections would prove beneficial in other entropy based avenues of computing. In future, we would try to hypothesize or assert why certain entropy measures work better with certain kinds of data and include more entropy measures for testing. We would also study the effect of using the generalized entropy measures in place of Shannon's entropy with other research problems, such as cluster validation, metric evaluation, and so on. .

\bibliographystyle{IEEEtran}
\nocite{*}
\bibliography{wcnps}

\end{document}